# Superconductivity in $Ba_2Sn_3Sb_6$ and $SrSn_3Sb_4$

**Laura Deakin,[a] Robert Lam,[a] Frank Marsiglio,[b] and Arthur Mar[a,*]**

[a] *Department of Chemistry, University of Alberta, Edmonton, AB Canada T6G 2G2*

[b] *Department of Physics, University of Alberta, Edmonton, AB Canada T6G 2J1*

[*] Corresponding author. Tel: (780) 492-5592. Fax: (780) 492-8231. E-mail: arthur.mar@ualberta.ca

**Abstract**

Resistivity and ac magnetic susceptibility measurements on $Ba_2Sn_3Sb_6$ and $SrSn_3Sb_4$ indicate that these Zintl compounds display a transition to a superconducting phase at $T_C$ = 3.9 K. The Meissner effect was observed for $Ba_2Sn_3Sb_6$ under an applied field of 25 Oe. The signatures for superconductivity, such as high and low velocity conduction electrons and lone pairs, are present for both of these compounds.

## 1. Introduction

Zintl compounds [1], consisting of an alkali or alkaline earth element combined with a p-block element, have generally been assumed to be semiconductors, despite the absence of experimental transport data in most cases. Where measurements have been made, the low-temperature data remain sparse. However, there is mounting evidence that challenges the assumption of semiconductivity, particularly in borderline Zintl compounds involving heavier p-block elements where the band gap can be narrowed considerably. For instance, some Zintl phases such as $Ba_8In_4Sb_{16}$ [2] and $BaGa_2Sb_2$ [3] are found to be narrow band gap p-type semiconductors, and a few such as $Ba_3Sn_4As_6$ are metallic [4]. Some binary borderline Zintl phases are superconducting, such as $BaSn_3$ ($T_C$ = 4.3 K)[5] and $SrSn_3$ ($T_C$ = 5.4 K) [6]. The observation of superconductivity has been connected to the simultaneous presence of both high and low velocity conduction electrons in the band structure [7], the existence of van Hove singularities (saddle points) near $E_F$ [8], and low dimensionality of the structure [9]. We aver that borderline metallic Zintl compounds are potentially a rich source of candidates displaying conventional superconductivity as both localized and itinerant electronic behaviour is often encountered, and their structures frequently display low-dimensional features in the anionic networks [10].

The ternary antimonides $Ba_2Sn_3Sb_6$ [11] and $SrSn_3Sb_4$ [12] are Zintl compounds that possess related channel structures made up of 30-membered rings in their Sn–Sb anionic frameworks (Figure 1). The channels run down the crystallographic *b* axis in both cases and accommodate the alkaline-earth cations. In $Ba_2Sn_3Sb_6$, zigzag Sb chains also reside within the channels. The resistivities were previously measured only down to 25 K and indicated metallic behaviour with an approach toward the origin in the case of $Ba_2Sn_3Sb_6$ [13]. This observation, along with characteristic signatures of flat and wide hole-like bands in the electronic band structure [13], prompted us to re-examine the physical properties at lower temperatures previously inaccessible to us. We present here evidence for superconductivity in the ternary antimonides $Ba_2Sn_3Sb_6$ and $SrSn_3Sb_4$.

## 2. Experimental

Single crystals of $Ba_2Sn_3Sb_6$ and $SrSn_3Sb_4$ were grown from Sn flux reactions as described previously [11, 12]. The products were characterized by powder X-ray diffraction patterns collected on an Enraf-Nonius FR552 Guinier camera (Cu $K\alpha_1$ radiation; Si standard) and by EDX (energy-dispersive X-ray) analysis on a Hitachi S-2700 scanning electron microscope.

Electrical resistivities were measured by standard four-probe techniques on a Quantum Design PPMS system equipped with an ac-transport controller (Model 7100). The current was 0.1 mA and the frequency was 16 Hz. The resistivity was measured along the needle axis (crystallographic *b* axis) of single crystals of $Ba_2Sn_3Sb_6$ and $SrSn_3Sb_4$. The compositions of all crystals used in these measurements were confirmed by EDX analysis. $Ba_2Sn_3Sb_6$: Anal. Calcd. (mol %) Ba, 18; Sn, 27; Sb, 55. Found: Ba, 14(2); Sn, 33(2); Sb, 53(2). $SrSn_3Sb_4$: Anal. Calcd. (mol %) Sr, 12; Sn, 38; Sb, 50. Found: Sr, 7(2); Sn, 41(2); Sb, 52(2). Critical temperatures were determined at the point at which the resistivity is 90% that of the normal value. The field dependence of the resistivity was measured upon warming the sample after it was cooled under zero-field conditions.

Magnetic measurements on $Ba_2Sn_3Sb_6$ were made with use of a Quantum Design PPMS 9T magnetometer/susceptometer. Ac susceptibility measurements under various dc fields were made upon warming the sample after it was cooled under zero-field conditions using a frequency of 1000 Hz and a driving amplitude of 1.0 Oe. The purity of the $Ba_2Sn_3Sb_6$ powder susceptibility sample was determined by Guinier powder X-ray diffraction. No lines characteristic of elemental Sn were observed. Any Sn impurities would be present at a level less than the detection limit of powder X-ray diffraction, on the order of less than 5%.

## 3. Results and discussion

The resistivity of single crystals of $Ba_2Sn_3Sb_6$ and $SrSn_3Sb_4$ down to 2 K is shown in Figure 2. There is a sudden decrease in resistivity at $T_C$ = 3.9 K for both compounds. With increasing applied field (insets in Figure 2), the critical temperatures shift systematically to lower values. At temperatures below $T_C$, there exists a residual resistivity, which is attributed to surface degradation of the crystals resulting from the treatment of 6M HCl used to remove excess Sn flux. A surface reaction forming an insulating layer would impede the resistivity from descending to zero.

The temperature dependence of the ac magnetic susceptibility in this low-temperature range confirmed superconductivity behaviour in $Ba_2Sn_3Sb_6$ (Figure 3a). The diamagnetic shielding (zero-field-cooled) and Meissner effect (field-cooled) were observed at 25 Oe, and these curves are superimposable. We have been unable to prepare a sufficiently pure sample of $SrSn_3Sb_4$ to measure its magnetic susceptibility. The values of $T_C$ for $Ba_2Sn_3Sb_6$ and $SrSn_3Sb_4$ do not coincide with that for elemental Sn ($T_C$ = 3.722 K) [14]. Measurements of $\chi'_{ac}$ for elemental Sn (Cerac, 99.8 %) and $Ba_2Sn_3Sb_6$ under identical instrumental conditions (Figure 3b) confirmed that $Ba_2Sn_3Sb_6$ displays a $T_C$ value that is noticeably different. Moreover, the use of single crystals, previously screened by EDX analyses, for the resistivity measurements precludes the possibility that the superconductivity arises from elemental Sn.

The appearance of superconductivity in $Ba_2Sn_3Sb_6$ and $SrSn_3Sb_4$ is consistent with the presence of disperse bands crossing the Fermi level and flat bands near the Fermi level in the band structure [13]. The disperse bands arise from orbital interactions oriented along the channel directions; the flat bands, perpendicular to them. The lone pairs associated with $SnSb_3$ trigonal pyramids in the crystal structures of both compounds may also be significant. There have been recent suggestions implicating a correlation between superconductivity and the presence of lone pairs in metallic conductors [5, 8] We note that in a previous study, the ternary antimonide $La_{13}Ga_8Sb_{21}$, containing pyramidally distorted $GaSb_3$ units, is a superconductor ($T_C$ = 2.4 K), whereas the related compound $La_{12}Ga_4Sb_{23}$, containing planar $GaSb_3$ units, is a normal metal [15]. Another suggestion is that the hole-like nature of the bands crossing the Fermi level may be significant for superconductivity [16, 17]. This theory can be tested by attempting to dope these compounds with holes and electrons, with the result that $T_C$ as a function of doping concentration will exhibit a familiar bell-shaped curve.

Low-temperature physical property measurements on $EuSn_3Sb_4$, isostructural with $SrSn_3Sb_4$ but containing unpaired 4*f* electrons, are currently being undertaken. More rigorous electronic structure calculations may also prove helpful.

**Acknowledgments**

The Natural Sciences and Engineering Research Council of Canada and the University of Alberta supported this work. We thank Wing Yan Chan for assistance with the preparation of $Ba_2Sn_3Sb_6$, and Christina Barker (Department of Chemical and Materials Engineering) for assistance with the EDX analyses.

**Figure captions**

Fig. 1. Structures of (a) $Ba_2Sn_3Sb_6$ and (b) $SrSn_3Sb_4$ viewed down the *b* axis. The large lightly-shaded circles are Ba or Sr atoms, the solid circles are Sn atoms, and the open circles are Sb atoms.

Fig. 2. Resistivity of (a) $Ba_2Sn_3Sb_6$ and (b) $SrSn_3Sb_4$. Insets show the field dependence of the resistivity between 0 and 500 Oe. Lines are included to guide the eye.

Fig. 3. (a) Field dependence of the zero-field-cooled ac magnetic susceptibility of $Ba_2Sn_3Sb_6$ and the field-cooled susceptibility acquired under a 25 Oe applied dc field. (b) Normalized ac magnetic susceptibilities of $Ba_2Sn_3Sb_6$ and elemental Sn under a zero dc field.

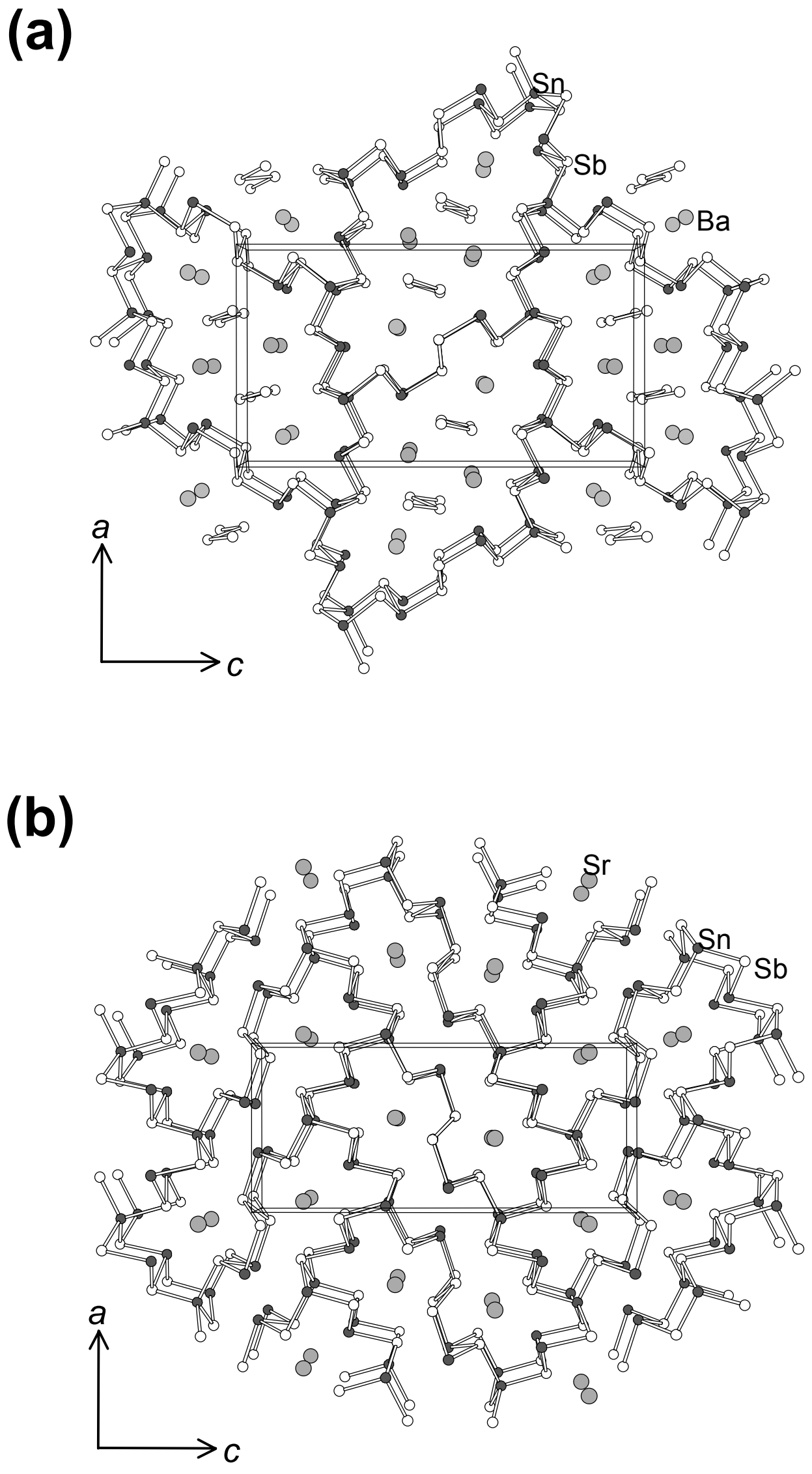

(a)

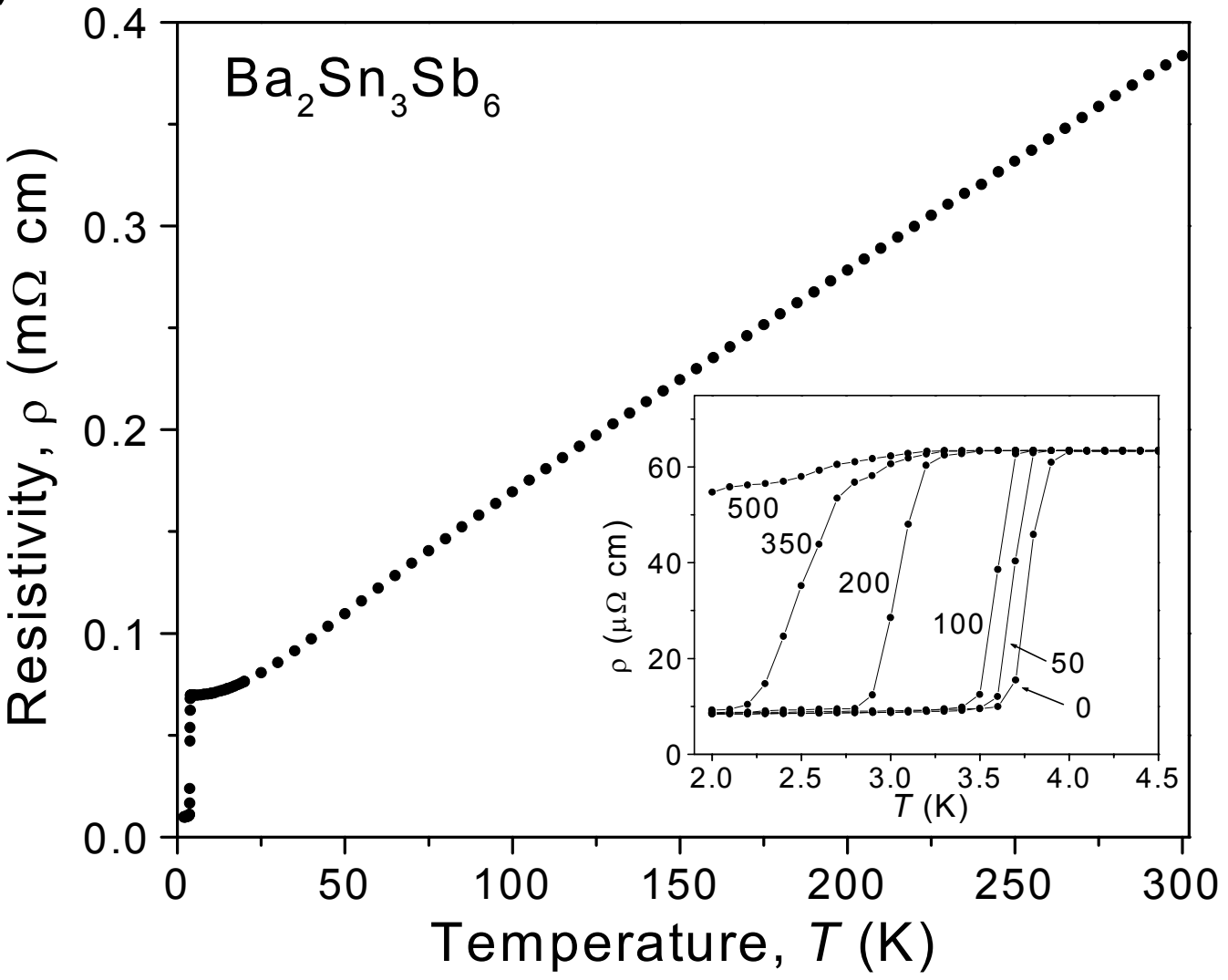

(b)

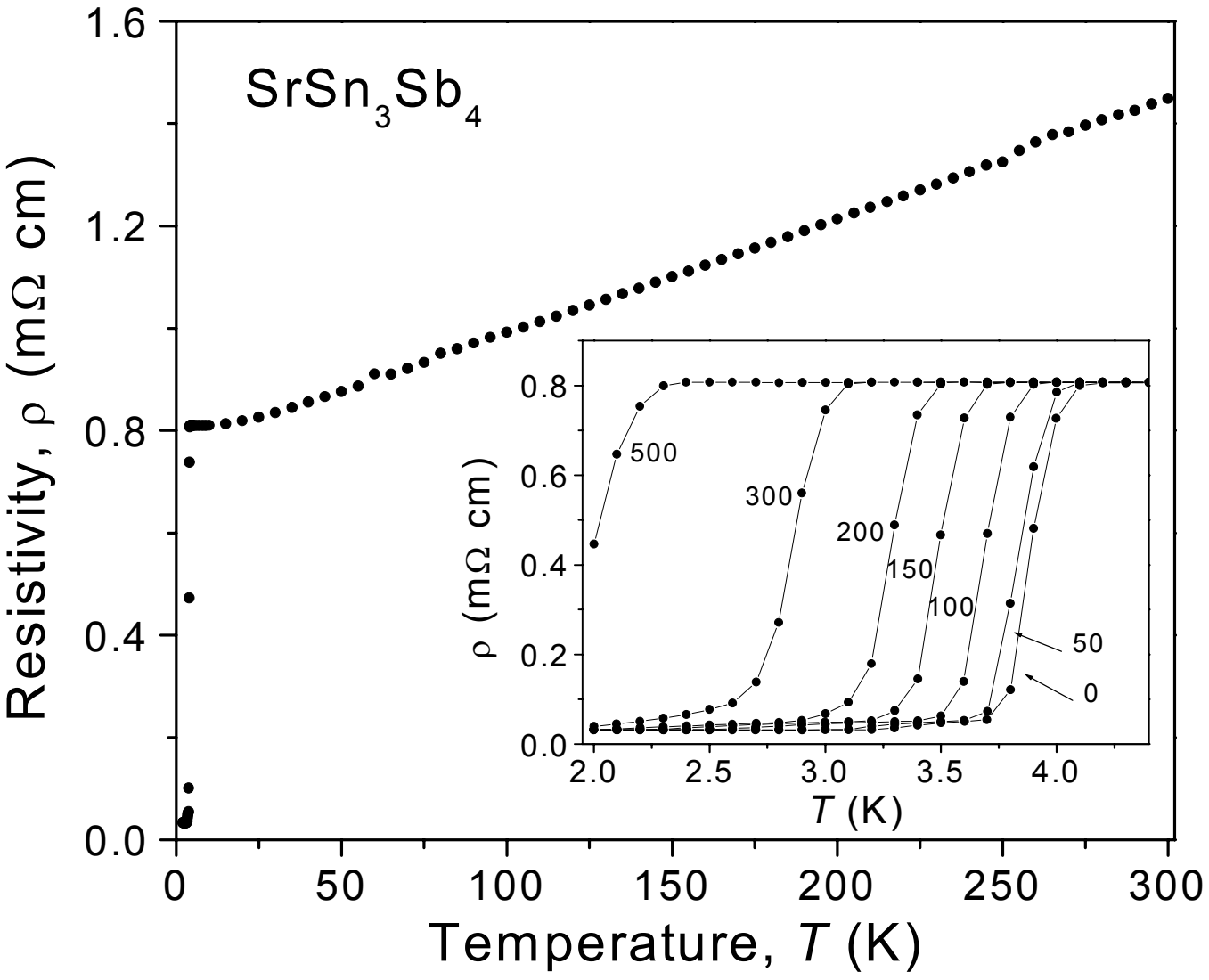

**(a)**

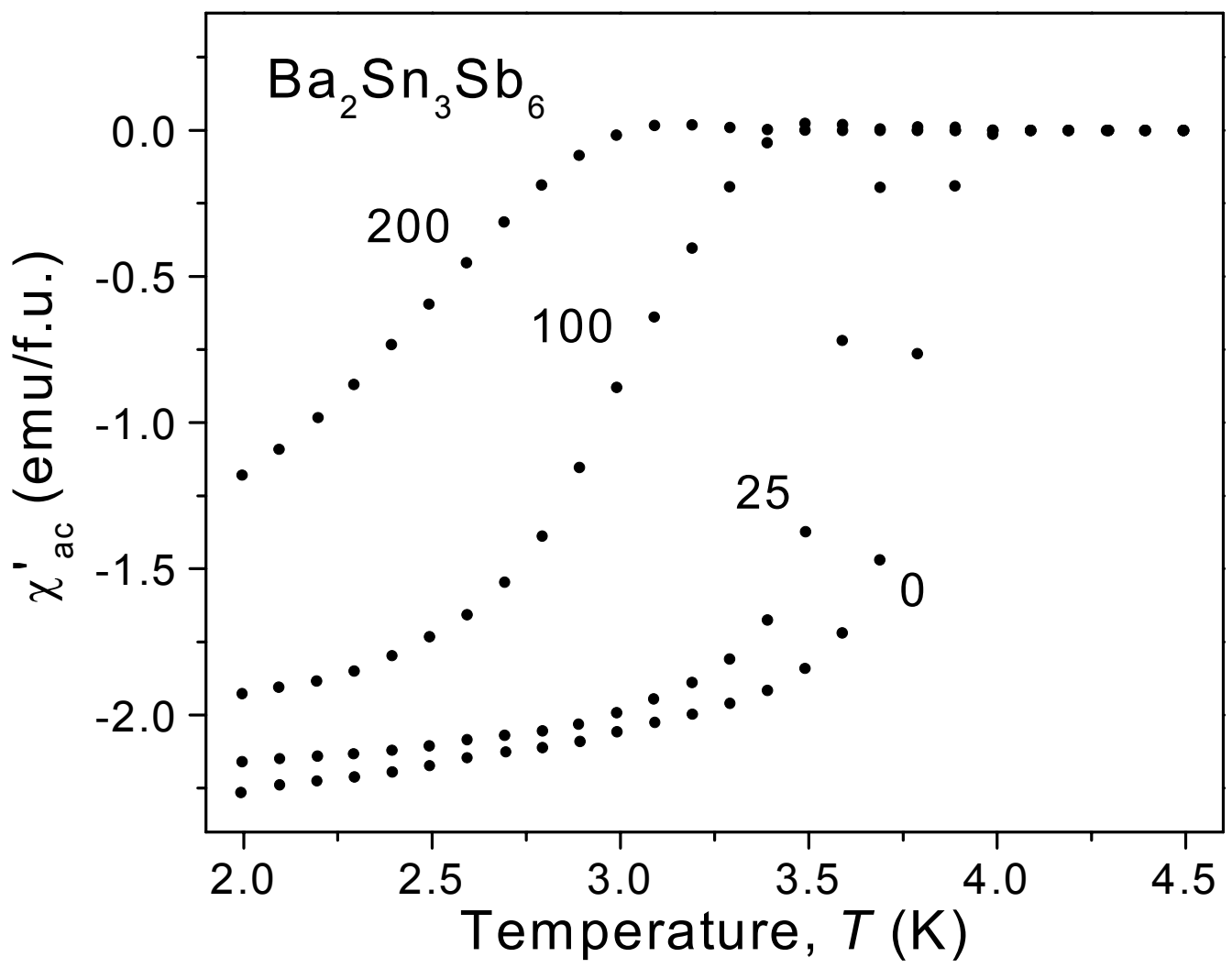

**(b)**

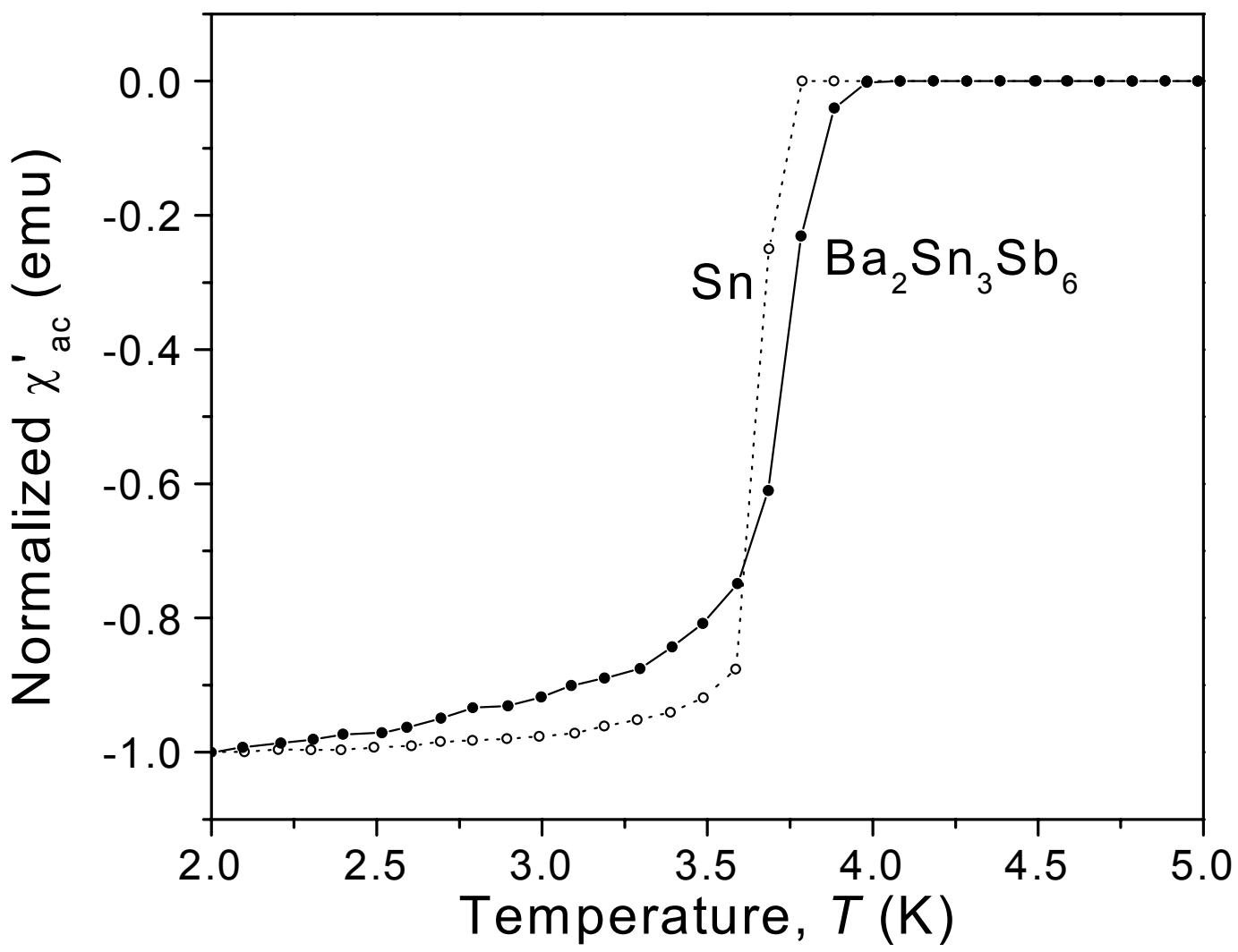
Sn
Ba2Sn3Sb6
Normalized χ'ac (emu)
Temperature, T (K)